\documentclass{PoS}

\title{ The INTEGRAL Galactic Plane Scanning}

\ShortTitle{Galactic Plane Scanning}

\author{{Mariateresa Fiocchi}\\
        Istituto di Astrofisica e Planetologia Spaziali. INAF. Italy\\
        E-mail: \email{mariateresa.fiocchi@iaps.inaf.it}}
\author{\speaker{Lorenzo Natalucci}\\
         Istituto di Astrofisica e Planetologia Spaziali. INAF. Italy\\
        E-mail: \email{lorenzo.natalucci@iaps.inaf.it}}
 \author{on behalf of the GPS Team\\
   }

\abstract{
After the first nine years of INTEGRAL operational life, the discovery of new sources
and source types, a large fraction of which are highly transient or highly absorbed, is certainly 
one of the most compelling results and legacies of INTEGRAL. 
Frequent monitoring of the Galactic Plane in AO8 and AO9 campaigns allowed us to detect transient sources, 
both known and new, confirming that the gamma-ray sky is dominated by the extreme variability of
different classes of objects. Regular scans of the Galactic Plane by INTEGRAL provide the most sensitive
hard X-ray wide survey to date of our Galaxy, with flux limits
of the order of 0.3 mCrab for an exposure time of $\sim$2Ms. Many transient sources have
been detected on a wide range of time scales ($\sim$~hours to months) and identified by triggered followup 
observations, mainly by Swift/XRT and optical/infrared telescopes.  
These discoveries are very important to characterize the X-ray binary population 
in our Galaxy, that is necessary input for evolution studies. 
The transient source monitoring is crucial to sample a variety of
physical conditions corresponding to a large span in luminosity. The long exposure times available
allow us to define the spectral and timing properties of the known sources and the global properties of each
object class. \\
}

\FullConference{"An INTEGRAL view of the high-energy sky (the first 10 years)"
9th INTEGRAL Workshop and celebration of the 10th anniversary of the launch,\\
		October 15-19, 2012\\
		Bibliotheque Nationale de France, Paris, France}

\begin{document}
\vspace{-0.2cm}
\section{The INTEGRAL Galactic Plane Scanning}
For the first 5 years of INTEGRAL operational life, the scientific Core Programme included 
regular scans of the Galactic Plane and Bulge. 
These observations led to the discovery of many new high energy sources. Many of these are  
the highly absorbed HMXB, that were previously unknown due to the lack of sensitive hard X-ray observations.
Furthermore, many new transient objects have been found, on short timescales and with low luminosity
($<10^{36} erg s^{-1}$).
From AO-5 onwards, these
regular scans were discontinued, resulting in a significant drop in the discovery rate of new Galactic systems. 
From AO8 to AO10 however, a multi-year proposal for Galactic Plane Scanning  
has been approved and since 2011 this program is collecting data for a total exposure time of $\sim$5~Msec (at the end of 2013)
to be added to the existing previous one, giving a total of $\sim135$~Ms.
The majority
of these sources are best detected on timescales less than or equal to an INTEGRAL orbit ($\sim$250ks), so that scans
occur once every revolution ($\sim$3 days), increasing the chances to detect short outbursts from unknown
and known sources and the possibility to monitor the evolution of long outbursts from
different source classes. The observed sky region cover a strip of
width $\pm$10 degree in galactic latitude and a longitude extent of about 170 deg. 
Due to its unique capability to monitor the Galactic Plane for long periods
at a sensitivity level unreachable by other wide-field imagers 
($<\sim0.3-0.5$~mCrab for T$_{exposure}$=1Ms, $\sim3$~mCrab for T$_{exposure}$=10ks), 
INTEGRAL has proven especially capable of discovering faint and/or short-lived phenomena that are 
typical of Galactic binary sources.   
\begin{figure}[h]
\centering
\includegraphics[scale=0.77]{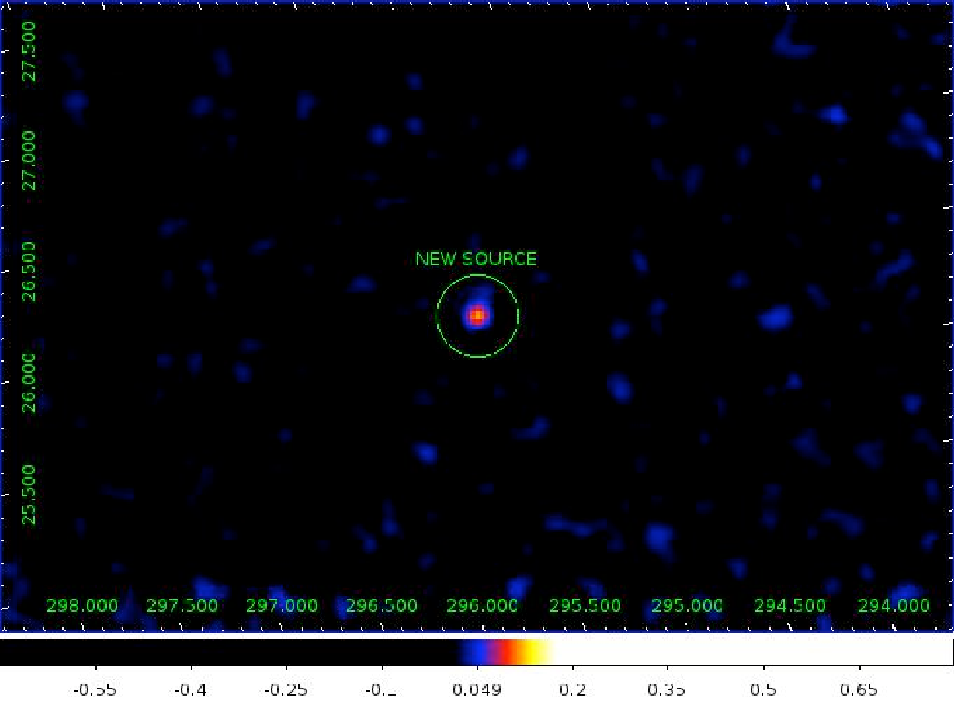}
\includegraphics[scale=0.82]{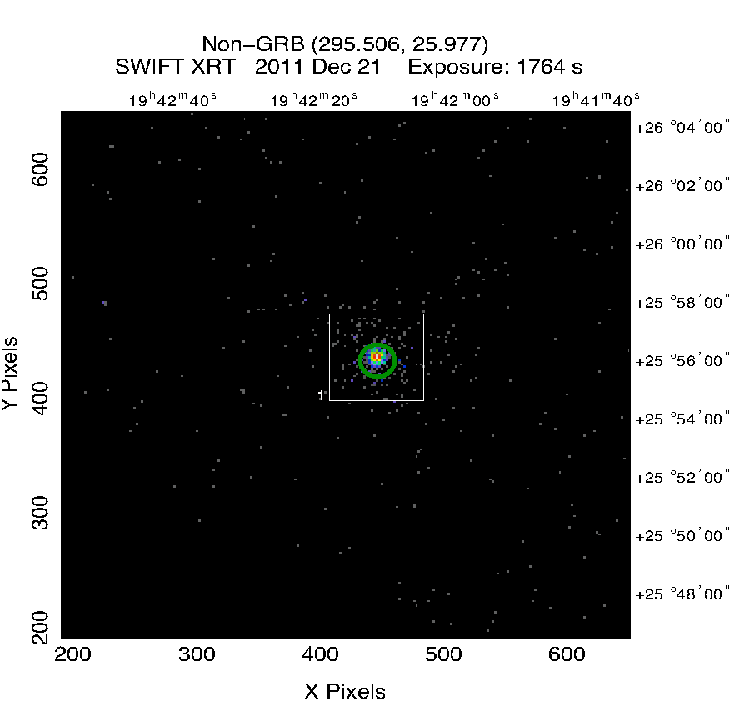}
\caption{Left: JEM-X image in the 3-10 keV image of 1RXS J194211.9+255552 . Right: The XRT/Swift image in the 2-10 kev of the 1RXS J194211.9+255552}
\label{fig:1}
\end{figure}

\begin{figure}[h]
\centering
\includegraphics[scale=0.30, angle=-90]{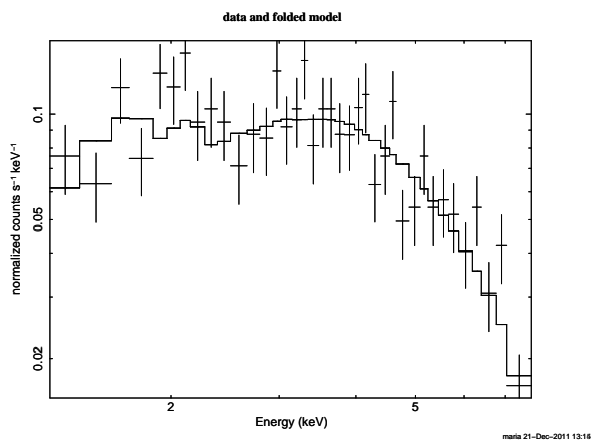}
\caption{The XRT spectrum of ROSAT  source 1RXS J194211.9+255552,  fitted with a power law model  with an absorption of N$_H\sim1.3\times10^{22} cm^{-2}$ and a photon index of $0.64\pm0.3$ }
\label{xrt}
\end{figure}
The Galactic Plane observations with INTEGRAL so far have changed our view of the Galaxy doubling the number
of known high-mass X-ray binaries (Bird et al. 2010), discovering
two new types of High Mass X-ray Binary (HMXB):  1) the high absorbed binary systems with N$_H\geq10^{24}cm^{-2}$, 
such that these sources could not be identified earlier with soft X-ray instruments (Kuulkers 2005;
Ibarra et al. 2007); 
2) the Super Giant Fast X-ray Transient (SFXT), showing the short (from tens of minutes to few hours) and intense flares (L$\sim10^{36-37} erg s^{-1}$), 
X-ray pulsar like spectrum, with very high dynamic range: 3-5 orders of magnitude with respect to the quiescent emission with $L\sim10^{32} erg s^{-1}$
(for review see Sidoli 2010, Sidoli 2011). 
The physics of these processes is still unknown and represent one of the main open issues in the study of compact objects in binary system.
Studies of these HMXBs classes will be fundamental to understand the population synthesis, to determine the chemical enrichment of the Galaxy and the evolution of massive stars in binaries and to investigate the poorly understood accretion mechanism (Sidoli 2009 for a review) which produces the short flares observed from the SFXT.

The gamma-ray satellites Fermi and AGILE and the 
Cherenkov telescopes (as HESS, Milagro etc.) have provided new surveys of the Galactic Plane from MeV to TeV energies.
INTEGRAL is and will be the unique opportunity to 
 identifying counterparts of the unidentified very high energy sources and observations
above 20 keV are needed to understand the physical origin of their emission.
 In the second Fermi/LAT catalogue about 320
MeV/GeV sources have been reported within 10 degrees of the Galactic Plane and about one third are still
unidentified (Nolan et al. 2013). Among them, some are transient sources flaring on timescales shorter than a few days, which can be a new
class of the  galactic fast transient high-energy emitters. 
 INTEGRAL represent a good possibility to systematically identify the unidentified MeV/GeV/TeV sources, define the spatial distributions and luminosity functions of the
each object classes.

The INTEGRAL observatory strategy and its large field of view allows to monitoring  a large number and variety of galactic sources 
for a long time, needed to study the physics of emission mechanisms for white dwarfs, neutron stars and black
holes.  These observations allowed us to understand the interplay between the different components of the accretion flow
onto the compact objects such as the cold disk and the neutron star, the hot Comptonizing plasma and the possible synchrotron
radiating jets  (Cadolle Bel et al., 2009; Prat et al., 2009; Fiocchi et al. 2008).//
In particular we report here some INTEGRAL result: 
1) non-thermal components at energies >100 keV from neutron star low-mass X-ray binaries (Fiocchi et
al. 2006; Paizis et al. 2009; Tarana et al. 2011);
2) INTEGRAL discovered high energy emission (200-300 keV) in a strong magnetized neutron
stars, with $B=10^{14-15}$ Gauss (Mereghetti et al. 2009 and and references therein)
and the Soft Gamma Repeaters spectral behavior,  soft at energies above 10 keV and hard below 10 keV, in
contrast to spectra from  Anomalous X-ray Pulsars  which show an opposite spectral shape;
3) hard emission from black holes up to 1 MeV (Caballero Garcia et al., 2007; Del Santo et al. 2012).

\vspace{-0.2cm}
\section{The Galactic Plane Scanning throughout AO8 and AO9}

We briefly report on the preliminary results from the recent AO8 and AO9 Programme, concerning the discovery of the new source, monitoring of the know source and the timing study on long time scale. The results reported here are simple examples of the many different topics that can be studied with Galactic Plane observations.

\subsection{The new detections}
At the time of writing, the pointings of the Vela and Cygnus region starting in May 2011, allowed us to
monitor many known sources: 31 HMXBs, 36 LMXBs,  5 AGN and 4 other sources (Atels
3361, 3434, 3816, 3818, 3887, 3916, 4135, 4136, 4168, 4218). 
The Galactic Plane Scanning observations (from AO1 to AO9) and the follow-up soft X-ray and optical/IR allowed to
identify new sources in the Galactic plane: $\sim$50 AGN hidden behind the galactic plane, $\sim$50 HMXB, $\sim$20 LMXB (Low Mass X-ray Binary)
and $\sim$20 CV (Cataclism Variables). 
Other $\sim$120 new IGR sources have been found but their highly variable behavior makes it difficult to identify them,
confirming that the gamma-ray sky is dominated by the extreme variability of different classes of objects (Bird et al. 2010).

A recent example of detection of a new source by INTEGRAL/JEM-X  happened during the Galactic Plane Scan observations performed between 2011 December 18, 13:47 UTC and December 19, 8:59 UTC. This source has been promptly identified as the transient ROSAT 1RXS J194211.9+255552, with averaged fluxes of 10$\pm$2 mCrab and 16$\pm$5 mCrab in the 3-10 keV and 10-25 keV energy ranges, respectively (Atel 3816).  
A Swift ToO follow-up has been performed on December 21, 2011 at 06:10:09.7 UTC, with a net exposure of 1756 s. 
Within the ROSAT error circle only one source is found 
at position 19h42m11.13s, +25:56:07.32 (J2000), 
with 3.6 arcsec error radius  (Atel 3818). Figure 1 shows the JEM-X image in the 3-10 keV energy range (left) and the XRT image in the 2-10 keV energy range (right), while in Figure~\ref{xrt} the XRT spectrum of 1RXS J194211.9+255552 is shown. The best fit consisting of a simple absorbed power law model with an absorption of N$_H=(1.3\pm0.6)\times10^{22} cm^{-2}$, a photon index of $0.7\pm0.3$ and an unabsorbed flux  of $\sim 7.9\times 10^{-11} erg cm^{-2} s^{-1}$, in 1-10 keV energy range (reduced $\chi^2$=1.1 for 36 degrees of freedom).   
Masetti et al. (Atel 4209) identified the optical counterpart using the 1.5m Cassini telescope of the Astronomical Observatory of Bologna in Loiano (Italy) 
and showed that the optical counterpart is a Galactic high mass X-ray binary, confirming the HMXB nature for this source as suggested by ATel 3818. 
\begin{figure}
\centering
\includegraphics[width=0.5\textwidth]{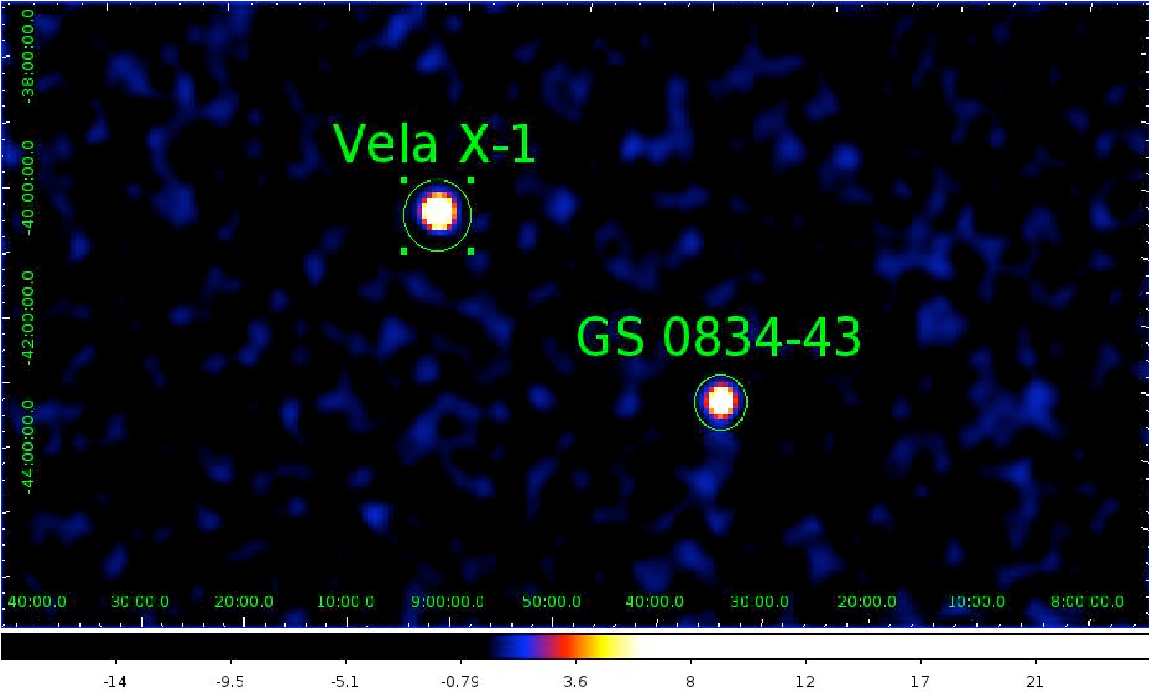}
\includegraphics[width=0.46\textwidth]{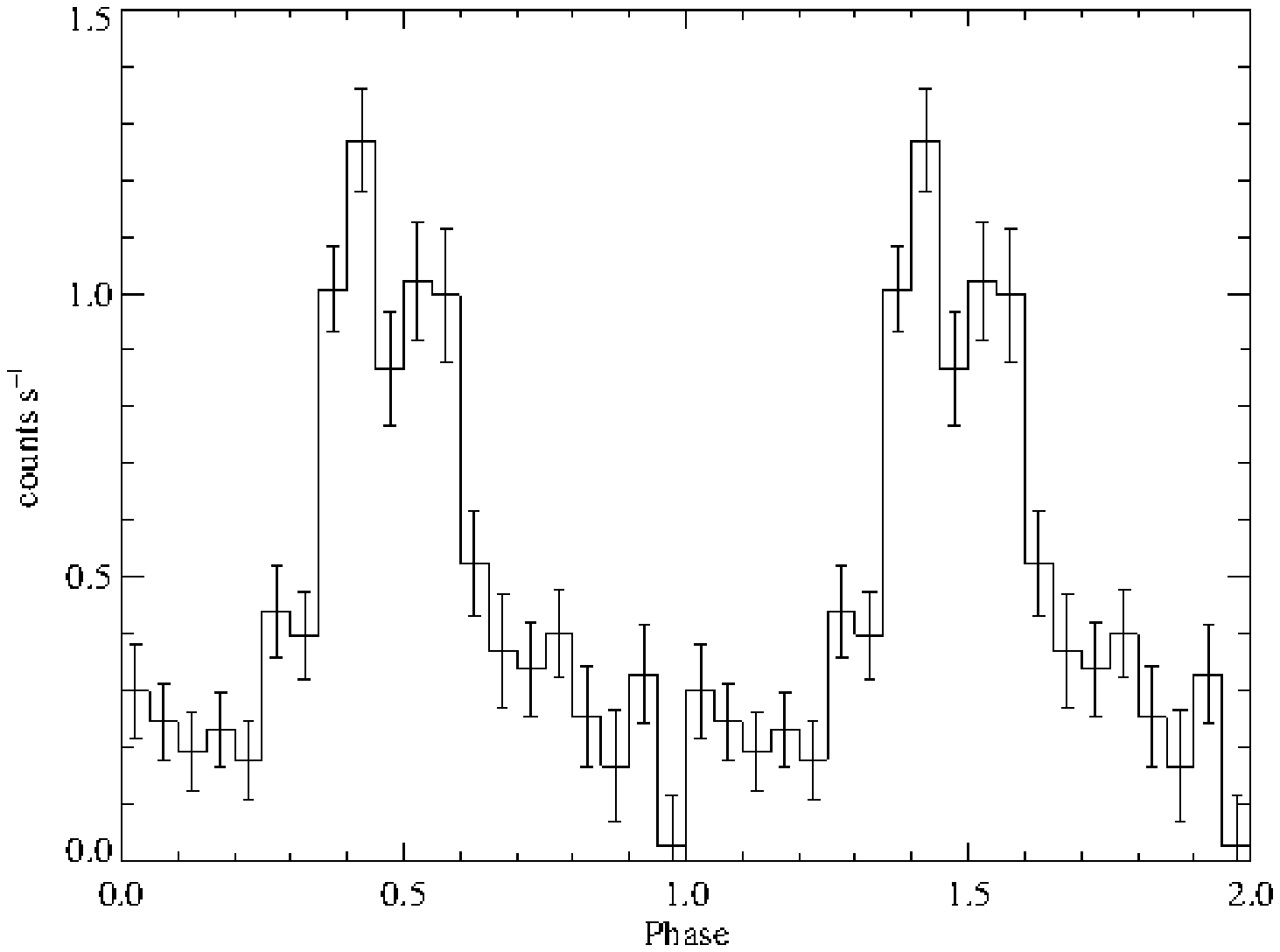}
\caption{Left: IBIS image in the 18-60 keV image during revolution 1184 of GS 0834-43. Right: The IBIS/ISGRI 18$-$60keV phase folded light curve using the best orbital period determination of 10.068 days of IGR J16328-4727 }
\label{fig:1}
\end{figure}

\subsection{The monitoring of known sources}

During the Galactic Plane Scanning INTEGRAL perform a long monitoring of GS~0834-43, a source discovered 
by Granat/WATCH in February 1990 with a peak flux of $\sim400$~mCrab in the 8-20 keV energy range (LapShov et al. 1992) and also observed at high energies by CGRO/BATSE at regular intervals until June 1993 (Wilson et al. 1997). 
These observations allowed to characterize GS~0834-43 as a Be transient X-ray pulsar with a spin period of $\sim$~12s, a wide orbit (period of $\sim$ 106 days), a small eccentricity between 0.10 and  0.17 and a low inclination angle (Wilson et al. 1997). 
After 1993, no outbursts were detected by BATSE or other monitoring instruments such as Swift/BAT (Cusumano et al. 2010) or INTEGRAL/IBIS (Bird et al. 2010).
During INTEGRAL Galactic Plane Scan observations between 2012-06-26 06:44:07 and 2012-06-26 15:27:19 UTC, IBIS/ISGRI detected this source at 47 sigma confidence level in the 18-60 keV energy range with a exposure time of 15.8 ks. It was also detected by the JEM-X  instrument at a significance of 19 sigma for an exposure time of 1.7 ks. The flux is $109\pm2$ mCrab in 18-60 keV and $64\pm4$ mCrab in the 3-10 keV energy range (Atel 4218). In Figure~3 (left) is shown the IBIS image of this field in the 18-60 keV energy range. This new, strong detection signature of a renewed period of activity after a long period of quiescence was also detected with FERMI/GRBM (Atel 4235). 
Throughout the duration of the observations, the IBIS light curve in the 18-60 keV energy range shows that the source flux remained constant.
The IBIS spectrum was well fitted with a simple power-law model with a photon index of $3.0\pm0.2$ and a flux of 
$1.8\times10^{-9} erg cm^{-2} s^{-1}$ ($\chi^2$=12/9 dof). 
 The INTEGRAL detectin of this new, strong outburst could be a signature of a renewed long term
activity period that is being followed up by other orbiting high energy telescopes, also covering the hard X-rays. The Nuclear Spectroscopic Telescope Array (Nustar) has also observed this source in July 2012 with two pointings and is expected to provide better sensitive spectral/timing measurements.  

\subsection{The timing studies}
The long exposure time reached in the Galactic Plane is an unique opportunity to perform timing studies, in particular to measure the orbital period of the X-ray binaries and to determine the spectral characteristics of the Galactic source.
The long term 18$-$60keV IBIS/ISGRI light curve of IGR J16328-4726 is a very meaningful example of the detection of a period at $\sim$10 days, which we interpret as the orbital period (for details see Fiocchi et al. 2012).  The folded light curve using the best orbital period determination of $\sim$10  days is shown in Figure~3 (right), where is detected a single broad emission maximum above a quiescent level with the shape similar to those observed from other SFXTs such as IGR~J17544-2619 (Clark et al. 2009) and IGR~J16465-4507 (Clark et al. 2010). 

\section{Conclusions}
Thanks to the combination of its large field of view and good sensitivity for short exposures (20 mCrab is reached at 50 keV with 2 standard GPS pointings of 2ksec each),  
timing resolution better than 1ms and good broad band spectroscopy,
INTEGRAL is and will be the only possibility to study very relevant topics as the new sources, the periodicity of Galactic sources, the spectral and temporal characteristic of the sources and of the different classes, their spatial distributions within our Galaxy and luminosity functions.
Because of the arc-minute location accuracy at energies above 20 keV, INTEGRAL is and will be unique to find the hard-X counterparts of new unidentified MeV/GeV/TeV sources, discovered with AGILE and Fermi or with the ground based Cerenkov telescopes. Finally its instruments, sensitive in the 3~keV to 10~MeV range, provide a unique link between the soft X-ray band covered by X-ray telescopes such as XMM-Newton or Chandra and the energy band of the high-energy gamma-ray space missions.
The future Galactic Plane observations will provide a more uniform and longer monitoring
of the known sources, new source detections and will also detect 
outburst or in peculiar spectral states, triggering dedicated ToO.


\vspace{-0.2cm}
\begin{thebibliography}{99} 
\scriptsize
\bibitem{1}  Bird et al. 2010, ApJS, 186,1  
\bibitem{2} Caballero Garcia et al., 2007, ApJ 669, 534
\bibitem{3} Cadolle Bel et al., 2009, A\&A 501, 1
\bibitem{4} Clark et~al., 2009, MNRAS, 399, 113
\bibitem{5} Clark et~al., 2010, MNRAS, 406, 75
\bibitem{6} Cusumano et al. 2010, A\&A, 510, 48
\bibitem{7} Del Santo et al. 2013, MNRAS, 430, 209
\bibitem{8} Fiocchi et al. 2006, ApJ, 651, 416
\bibitem{9} Fiocchi et al. 2008, A\&A,492,557
\bibitem{10} Fiocchi et al. 2012, ApJ, 762, 19
\bibitem{11} Kuulkers 2005, AIPC News 797, 402
\bibitem{12} Ibarra et al., 2007, A\&A ,465, 501
\bibitem{13} LapShov et al. 1995, Soviet Astronomy Letters, Vol. 18, p. 12
\bibitem{14} Mereghetti et al. 2009, ApJ 696, L74
\bibitem{15} Nolan P.L. et al., 2012, ApJS, 199, 31
\bibitem{16} Paizis et al. 2009, PASJ, 61, 107
\bibitem{17} Prat et al., 2009, A\&A 494, L21
\bibitem{18} Sidoli 2009, Advances in Space Research, 43, 1464
\bibitem{19} Sidoli 2010, Symposium E, session 16, paper number E16-0026-10, COSPAR 38.2556S
\bibitem{20} Sidoli et al. 2011, astro-ph/1111.5747
\bibitem{21} Tarana et al. 2011, MNRAS, 416, 873
\bibitem{22} Wilson et al. 1997, ApJ, 479, 388
\end{thebibliography}
\end{document}